\documentstyle[12pt]{article}

\newcommand{\be}{\begin{equation}}
\newcommand{\ee}{\end{equation}}

\begin{document}

\vspace{1cm}
\begin{flushright}
CAMS/99-04\\
hep-th/9911180
\end{flushright}
\vspace{1cm} \baselineskip=16pt

\begin{center}
\baselineskip=16pt
\centerline{{\Large{\bf $D=7$ $SU(2)$ Gauged
Supergravity}}} \centerline{{\Large{\bf From $D=10$ Supergravity}}}

\vskip1 cm

A. H. Chamseddine and W. A. Sabra \vskip1cm
\centerline{\em Center
for Advanced Mathematical Sciences (CAMS), } \centerline{\em and}
\centerline{\em Physics Department, American University of Beirut,
Lebanon}
\end{center}

\vskip1 cm \centerline{\bf ABSTRACT} The theory of $SU(2)$ gauged
seven-dimensional supergravity is obtained by compactifying ten dimensional $%
N=1$ supergravity on the group manifold $SU(2)$. \vfill\eject
\bigskip

\section{Introduction}

Theories of extended supergravity in various dimensions possess rigid
symmetries. A subgroup of these symmetries can be gauged by the vector
fields present in the theory. For example, in $N=2$ supergravity in five
dimensions, one can gauge the $U(1)$ subgroup of the $SU(2)$ rigid symmetry
group of the theory \cite{cremmer81}. Gauged supergravity theories exist in
higher dimensions in which supersymmetry allows the existence of a
cosmological constant. It is well known that a cosmological constant is not
allowed in $d=11$, $d=10$ and $d=9$. In $d=7$, $SU(2)$ and $SO(5)\times
SO(5) $ gauged supergravity theories were constructed in \cite{TV} and \cite
{seven} respectively.

The theories of gauged supergravity theories in $d=4,5,7$ are believed to be
related to certain compactifications of $d=10,11$ supergravity theories. For
instance, the four dimensional gauged $N=8$ supergravity \cite{DN} was
conjectured to be related to the compactification of the original eleven
dimensional supergravity \cite{CJS} on $S^{7}$ \cite{DNP}. This conjecture
was proved by de Wit and Nicolai \cite{DN87} \ for a different formulation
of eleven dimensional supergravity with local $SU(8)$ invariance \cite{DN86}%
. The connection to the non linear Kaluza-Klein ansatz of the original $d=11$
Lagrangian could only be solved in certain sectors. Toroidal
compactification of ten dimensional supergravity to four dimensions yields
an $N=4$ supergravity theory with six vector multiplets \cite{chams1}. The
vector and matter fields obtained in four dimensions are in general linear
combinations of the internal components of the ten dimensional metric and
antisymmetric tensor. The truncation is then performed by identifying the
vector components of the ten-dimensional metric with those of the
antisymmetric tensor. Another known compactifications of ten dimensional
supergravity are due to Scherk and Schwarz \cite{SS}, in which the internal
compactified space is taken to be a group manifold. The maximal group
manifold allowed is $S^{3}\times S^{3}$ and a compactification of this
particular case for the dual formulation of supergravity was performed in
\cite{chams2}. The resulting four dimensional theory is an $N=4$
supergravity with a non-compact gauge group containing the factor $%
SU(2)\times SU(2)$. In order to obtain the Friedman-Schwarz model \cite{FS},
a compactification of the Sherck-Schwarz type is needed. This
compactification was performed recently in \cite{CV}. The crucial point in
this analysis is the specific relation between the components of the metric
and antisymmetric tensor along the internal dimensions. This was a long
standing problem, the main difficulty was finding the right ansatz for the
antisymmetric tensor field. It was suggested by the authors of \cite{DTN}
that a more complicated six dimensional internal manifold is needed in order
to obtain the Friedman-Schwarz model. This suggestion was motivated by the
general result of Friedman, Gibbons and West \cite{FGW} that non-trivial
compactifications of ten-dimensional supergravity are inconsistent. This
proved not to be the case as a basic assumption made in \cite{FGW} regarding
the dilaton field needs to be violated.

In recent years, there has been a renewed interest in gauged supergravity
theories. This is mainly due to the recently conjectured duality between
supergravity and super Yang Mills \cite{ads}. Since this conjecture has been
made, anti-de Sitter spaces have received a great deal of interest. The
purpose of our work here is to demonstrate that $d=7$, $N=2,$ $SU(2)$-gauged
supergravity theory of \cite{TV} can be obtained via dimensional reduction
of $N=1$ ten dimensional and eleven dimensional supergravity theories.

\section{Dimensional Reduction and $D=7$ Gauged Supergravity}

In this section we will show explicitly how to obtain the seven dimensional
gauged supergravity of \cite{TV} as a dimensionally reduced ten dimensional $%
N=1$ supergravity theory. The internal space is taken to be the group
manifold $SU(2)$.

The bosonic part of $N=1$ supergravity action in ten dimensions is
\begin{eqnarray}
S_{10} &=&\int \left( -\frac{1}{4}\,\hat{R}+\frac{1}{2}\,\partial _{M}\hat{%
\phi}\,\partial ^{M}\hat{\phi}+\frac{1}{12}\,e^{-2\hat{\phi}}\hat{H}_{MNP}\,%
\hat{H}^{MNP}\right) \,d^{4}x\,d^{6}z  \nonumber \\
&\equiv &S_{\hat{G}}+S_{\hat{\phi}}+S_{\hat{H}}.  \label{tenaction}
\end{eqnarray}
Our notations are as follows. Ten-dimensional quantities are denoted by
hatted symbols. Base space and tangent space indices are denoted by late and
early capital Latin letters, respectively. We first comment briefly on the
compactification of ten-dimensional supergravity on $S^{3}\times S^{3}$ to
four dimensions.

For four-dimensional space-time indices, late and early Greek letters denote
base space and tangent space indices, respectively. Similarly, the internal
base space and tangent space indices are denoted by late and early Latin
letters, respectively.
\begin{equation}
\{M\}=\{\mu =0,\ldots ,3;\,m=1,\ldots ,6\},\ \ \ \{A\}=\{\alpha =0,\ldots
,3;\,a=1,\ldots ,6\}.\ \
\end{equation}
The general coordinates $\hat{x}^{M}$ consist of spacetime coordinates $x^{{%
\mu }}$ and internal coordinates $z^{m}$. The flat Lorentz metric of the
tangent space is chosen to be $(+,-,\ldots ,-)$ with the internal dimensions
all spacelike. Thus the metric is related to the vielbein by
\begin{equation}
\hat{{\bf g}}_{MN}=\hat{\eta}_{AB}\hat{e}_{\ M}^{A}\hat{e}_{\ N}^{B}=\eta
_{\alpha \beta }\hat{e}_{\ M}^{\alpha }\hat{e}_{\ N}^{\beta }-\delta _{ab}%
\hat{e}_{\ M}^{a}\hat{e}_{\ N}^{b}\,,
\end{equation}
and the antisymmetric tensor field strength is
\begin{equation}
\hat{H}_{MNP}=\partial _{M}\hat{B}_{NP}+\partial _{N}\hat{B}_{PM}+\partial
_{P}\hat{B}_{MN}\,.
\end{equation}
The coordinates $z^{m}$ span the internal compact group space, implying that
we have the functions $U_{m}^{a}(z)$ satisfying the condition
\begin{equation}
\left. \left. \left( U^{-1}\right) _{a}^{\ m}\left( U^{-1}\right) _{b}^{\
n}\right( \partial _{m}U_{\ n}^{c}-\partial _{n}U_{\ m}^{c}\right) =\frac{%
f_{abc}}{\sqrt{2}}\,,
\end{equation}
Here $f_{abc}$ are the group structure constants and the internal space
volume is $\Omega =\int |U_{m}^{a}|d{\bf z}$. In the maximal case, i. e., $%
SU(2)\times SU(2)$, each $S^{3}$ factor admits invariant 1-form $\theta
^{a}=\theta _{i}^{a}dz^{i}$, which satisfies
\begin{equation}
d\theta ^{a}+{\frac{1}{2}}\epsilon _{abc}\theta ^{b}\wedge \theta ^{c}=0
\end{equation}
If one chooses
\begin{equation}
U_{m}^{a}\equiv U_{i}^{a}=-{\frac{\sqrt{2}}{g}}\theta _{i}^{a}
\end{equation}
where $g$ is a coupling constant, then the structure constants will be given
in terms of the coupling constant by $f_{abc}=g\varepsilon _{abc}$. For the
case where the coupling constant of one of the $SU(2)$ factors vanishes, the
internal space becomes the group manifold $SU(2)\times \lbrack U(1)]^{3}$.
In order to get a consistent truncation, one must set six vector multiplets
to zero. This is done by identifying the six gauge fields coming from the
components of the metric with six vector fields from the components of the
antisymmetric tensor. In order to do this identification the vectors coming
from the antisymmetric tensor must behave like Yang-Mills gauge fields, and
here is the main source of difficulty. It is crucial to have the right
ansatz for the antisymmetric tensor to get a consistent truncation.

We now turn to the compactification of ten-dimensional supergravity theory
down to seven dimensions. The three dimensional internal space is taken to
be the $SU(2)$ group manifold. We shall show that the obtained theory is the
$SU(2)$ gauged seven dimensional supergravity theory derived in \cite{TV}.
Following Scherk and Schwarz \cite{SS}, we parameterize the $vielbein$ in
the following form
\begin{equation}
\widehat{e}_{M}^{A}=\left(
\begin{array}{cc}
e^{\frac{3}{10}\hat{\phi}}e_{\mu }^{\alpha }(x) & \sqrt{2}e^{-\frac{1}{2}%
\hat{\phi}}A_{\mu }^{m}\left( z\right) \delta _{m}^{a} \\
0 & e^{-\frac{1}{2}\hat{\phi}}U_{m}^{a}\left( z\right)
\end{array}
\right)
\end{equation}
where \ we have set $\kappa =1$ and rescaled the gauge fields by ${\frac{1}{%
\sqrt{2}}}$. Here $U$ depends on the internal coordinates $(7,8,9)$ and $%
\hat{\phi}=\hat{\phi}\left( x\right) $.

Our ansatz in terms of the metric components is thus given by
\begin{eqnarray}
{\hat{g}}_{\mu \nu } &=&e^{{\frac{3}{5}}\hat{\phi}}g_{\mu \nu }-2A_{\mu
}^{a}A_{\nu }^{a}e^{-\hat{\phi}}  \nonumber \\
{\hat{g}}_{\mu m} &=&-\sqrt{2}U_{m}^{a}A_{\mu }^{a}e^{-\hat{\phi}}  \nonumber
\\
{\hat{g}}_{mn} &=&-U_{m}^{a}U_{n}^{a}e^{-\hat{\phi}}
\end{eqnarray}

Using equation (38) of \cite{SS},\footnote{%
When using this formula, we rescale the gauge fields} for the reduction of $%
S_{\hat{G}}$, one obtains the reduced Lagrangian, which reads
\begin{equation}
{\cal L}_{G}=-{\frac{1}{4}}R-{\frac{1}{8}}e^{-{\frac{8}{5}}\hat{\phi}}F_{\mu
\nu }^{a}F^{\mu \nu a}+{\frac{3}{10}}g^{\mu \nu }\partial _{\mu }\hat{\phi}%
\partial _{\nu }\hat{\phi}+{\frac{3g^{2}}{16}}e^{{\frac{8}{5}}\hat{\phi}}
\end{equation}
where
\begin{equation}
F_{\mu \nu }^{a}=\partial _{\mu }A_{\nu }^{a}-\partial _{\nu }A_{\mu
}^{a}+f_{abc}A_{\mu }^{b}A_{\nu }^{c}
\end{equation}
For the antisymmetric tensor our ansatz is
\begin{equation}
{\hat{B}}_{\mu \nu }={B}_{\mu \nu },\quad {\hat{B}}_{\mu m}=-{\frac{1}{\sqrt{%
2}}}{A}_{\mu }^{a}U_{m}^{a},\quad {\hat{B}}_{mn}={\tilde{B}}_{mn}
\end{equation}
where ${B}_{\mu \nu }$ is a function of $x$ and ${\tilde{B}}_{mn}$ is a
function of the internal coordinates $z$ only. The field strengths are given
by
\begin{eqnarray}
\hat{H}_{\mu \nu \rho } &=&H_{\mu \nu \rho }\equiv \partial _{\mu }B_{\nu
\rho }+\partial _{\nu }B_{\rho \mu }+\partial _{\rho }B_{\mu \nu }  \nonumber
\\
\hat{H}_{\mu \nu m} &=&-{\frac{1}{\sqrt{2}}}(\partial _{\mu }A_{\nu
}^{a}-\partial _{\nu }A_{\mu }^{a})U_{m}^{a}  \nonumber \\
\hat{H}_{\mu mn} &=&{\frac{1}{2}}f_{abc}A_{\mu }^{a}U_{m}^{b}U_{n}^{c}
\nonumber \\
\hat{H}_{mnp} &=&\partial _{m}{\tilde{B}}_{np}+\partial _{n}{\tilde{B}}%
_{pm}+\partial _{p}{\tilde{B}}_{mn}
\end{eqnarray}
where we also require that
\begin{equation}
\hat{H}_{mnp}={\frac{1}{2\sqrt{2}}}f_{abc}U_{m}^{a}U_{n}^{b}U_{p}^{c}
\end{equation}

The reduction of $S_{\hat{B}}$ gives the following
\begin{equation}
{\cal L}_{B}=\left( {\frac{1}{12}}e^{-{\frac{16}{5}}\hat{\phi}}H_{\mu \nu
\rho }^{\prime }H^{\prime \mu \nu \rho }-{\frac{1}{8}}e^{-{\frac{8}{5}}\hat{%
\phi}}F_{\mu \nu }^{a}F^{\mu \nu a}-{\frac{g^{2}}{16}}e^{{\frac{8}{5}}\hat{%
\phi}}\right)  \label{Hreduction}
\end{equation}
where
\begin{eqnarray}
H_{\mu \nu \rho }^{\prime } &=&H_{\mu \nu \rho }-\omega _{\mu \nu \rho }
\nonumber \\
\omega _{\mu \nu \rho } &=&-6(A_{[\mu }^{a}\partial _{\nu }A_{\rho ]}^{a}+{%
\frac{1}{3}}f_{abc}A_{\mu }^{a}A_{\nu }^{b}A_{\rho }^{c})
\end{eqnarray}
In deriving equation $\left( \ref{Hreduction}\right) $ care must be taken to
include the off-diagonal components of the metric. This can be effectively
done by first defining $\hat{H}_{ABC}=e_{A}^{M}e_{B}^{N}e_{C}^{P}\hat{H}%
_{MNP}$ and then writing
\begin{eqnarray*}
\hat{H}_{\alpha \beta \gamma } &=&e^{-\frac{9}{10}\hat{\phi}}e_{\alpha
}^{\mu }e_{\beta }^{\nu }e_{\gamma }^{\rho }H_{\mu \nu \rho }^{\prime }, \\
\hat{H}_{\alpha \beta \gamma } &=&e^{-\frac{1}{10}\hat{\phi}}e_{\alpha
}^{\mu }e_{\beta }^{\nu }F_{\mu \nu }^{a}, \\
\hat{H}_{\alpha \beta \gamma } &=&0, \\
\hat{H}_{abc} &=&e^{\frac{3}{2}\hat{\phi}}\frac{g}{2\sqrt{2}}\epsilon _{abc}.
\end{eqnarray*}
The appearance of the Chern-Simons term $\omega _{\mu \nu \rho }$ in $H_{\mu
\nu \rho }^{\prime }$ is a crucial test of the consistency of the ansatz.

The reduction of the scalar part of ten dimensional supergravity, $S_{\hat{%
\phi}}$ gives the following contribution to the seven dimensional Lagrangian
\begin{equation}
{\cal L}_{S}={\frac{1}{2}}\partial _{\mu }\hat{\phi}\partial ^{\mu }\hat{\phi%
}
\end{equation}
Therefore, combining ${\cal L}_{G}$, ${\cal L}_{B}$ and ${\cal L}_{S}$, the
seven dimensional theory is described by the Lagrangian
\begin{equation}
{\cal L}_{7}=-{\frac{1}{4}}R-{\frac{1}{4}}e^{-{\frac{8}{5}}\hat{\phi}}F_{\mu
\nu }^{a}F^{\mu \nu a}+{\frac{4}{5}}g^{\mu \nu }\partial _{\mu }\hat{\phi}%
\partial _{\nu }\hat{\phi}+{\frac{1}{12}}e^{-{\frac{16}{5}}\hat{\phi}}H_{\mu
\nu \rho }^{\prime }H^{^{\prime }\mu \nu \rho }+{\frac{g^{2}}{8}}e^{{\frac{8%
}{5}}\hat{\phi}}  \label{sevenL}
\end{equation}

The $N=2$ $SU(2)$ gauged $d=7$ supergravity which was constructed in \cite
{TV} is given by
\begin{eqnarray}
e^{-1}{\cal L} &=&-{\frac{1}{2}}R-{\frac{e}{48}}\sigma ^{-4}F_{\mu \nu \rho
\sigma }F^{\mu \nu \rho \sigma }-{\frac{\sigma ^{2}}{4}}F_{\mu \nu i}^{\quad
\,\,\,j}F_{\quad j}^{\mu \nu i}-{\frac{1}{2}}\partial _{\mu }\phi \partial
^{\mu }\phi  \nonumber \\
&&+{\frac{i}{48\sqrt{2}}}\epsilon ^{\mu \nu \rho \sigma \kappa \lambda \eta
}F_{\mu \nu \rho \sigma }F_{\kappa \lambda i}^{\quad j}A_{\eta j}^{\quad i}+{%
\alpha ^{2}}\sigma ^{-2}  \label{TV}
\end{eqnarray}
where $\sigma =e^{-\frac{1}{\sqrt{5}}\phi }.$ The Lagrangian in $\left( \ref
{sevenL}\right) $ can be seen to be identical to $\left( \ref{TV}\right) $,
after multiplying equation $\left( \ref{sevenL}\right) $ by an overall
factor of 2 and under the following identifications
\begin{eqnarray}
\hat{\phi} &=&{{\frac{\sqrt{5}}{4}}\phi ,} \\
g &=&2\alpha , \\
A_{\mu i}^{\quad j} &=&-A_{\mu }^{a}\left( \tau ^{a}\right)_{i}^{j}.
\end{eqnarray}

The Lagrangian $\left( \ref{TV}\right) $ contains a three-form $A_{\mu \nu
\rho }$ instead of the two-form $B_{\mu \nu }$ appearing in $\left( \ref
{sevenL}\right) $. These forms are, however, related by a duality
transformation. To see this we add to the seven dimensional compactified
action in $\left( \ref{sevenL}\right) $ the term
\[
-\frac{1}{36}\int d^{7}x\epsilon ^{\mu \nu \rho \sigma \kappa \lambda \eta
}H_{\mu \nu \rho }\partial _{\sigma }A_{\kappa \lambda \eta }
\]
and assume that $H_{\mu \nu \rho }$ is not the field strength of $B_{\mu \nu
}$ but an independent field. The equation of motion of $A_{\kappa \lambda
\eta }$ then implies that $H_{\mu \nu \rho }$ is the field strength of a two
form $B_{\mu \nu }$. On the other hand, now the $H_{\mu \nu \rho }$ appears
quadratically and linearly in the Lagrangian, the Gaussian integration of $%
H_{\mu \nu \rho }$ can be carried out resulting in the terms
\[
\int d^{7}x\left( \frac{e}{96}e^{\frac{16}{5}\hat{\phi}}F_{\mu \nu \rho
\sigma }F^{\mu \nu \rho \sigma }-\frac{1}{144\sqrt{2}}\epsilon ^{\mu \nu
\rho \sigma \kappa \lambda \eta }F_{\mu \nu \rho \sigma }\omega _{\kappa
\lambda \eta }\right) .
\]
Integrating the last term by parts and using the identity
\[
\partial _{\left[ \mu \right. }\omega _{\nu \rho \sigma \left. {}\right] }=-%
\frac{3}{2}F_{\mu \nu }^{a}F_{\rho \sigma }^{a},
\]
one obtains the Chern-Simons term
\[
\frac{1}{24\sqrt{2}}\int d^{7}x\epsilon ^{\mu \nu \rho \sigma \kappa \lambda
\eta }F_{\mu \nu }^{a}F_{\rho \sigma }^{a}A_{\kappa \lambda \eta }
\]
and this is seen to agree with the Lagrangian in $\left( \ref{TV}\right) $
after integrating by parts\footnote{%
The different signs for the kinetic terms as well as the i factor appearing
with the epsilon tensor in $\left( \ref{TV}\right) $ are due to the choice
of the metric $\left( +++++++\right) $ in \cite{TV}}.

We note in passing that the seven dimensional Lagrangian can also be
obtained by compactifying eleven dimensional supergravity and then
truncating. This can be seen by embedding the ten-dimensional supergravity
into eleven-dimensional supergravity after truncating half the degrees of
freedom. One has the following identifications for the eleven dimensional
fields:
\begin{eqnarray*}
e_{M}^{A} &=&e^{-\frac{1}{6}\hat{\phi}}\hat{e}_{M}^{A} \\
e_{\stackrel{.}{11}}^{11} &=&e^{\frac{3}{4}\hat{\phi}} \\
e_{M}^{11} &=&0 \\
e_{\stackrel{.}{11}}^{A} &=&0 \\
A_{MN\stackrel{.}{11}} &=&\hat{B}_{MN} \\
A_{MNP} &=&0
\end{eqnarray*}
This identification is important in lifting special solutions from seven to
ten and eleven dimensions.

We now comment on related work that appeared in the literature. First in the
work of Duff, Townsend and van Niewenhuizen \cite{DTN} a compactification of
ten dimensional supergravity on $S^{3}$ was given, but this breaks all
supersymmetries. Recently, a consistent truncation of eleven dimensional
supergravity to $N=4$ seven dimensional supergravity has been given by
Nastase, Vaman and van Niewenhuizen \cite{NVV}. They gave the complete non
linear Kaluza-Klein reduction on $AdS_{7}\times S_{4}$. Lu and Pope \cite{LP}
gave an ansatz for the reduction and truncation of eleven dimensional
supergravity to seven dimensions. The Lagrangian they obtained has two
coupling constants $g$ and $m$ where $m$ is the coefficient of the
topological term
\[
m\epsilon ^{\mu \nu \rho \sigma \kappa \lambda \eta }F_{\mu \nu \rho \sigma
}A_{\kappa \lambda \eta }.
\]
which is only possible in the $A_{\mu \nu \rho }$ formulation of $d=7$
supergravity but not in the $A_{\mu \nu }$ formulation. The case considered
in this paper corresponds to setting $m=0$ in \cite{LP} and is only obtained
from their work as a singular limit.

In this work we have not given the ansatz for the reduction of the fermionic
parts. This should be straightforward but tedious. The consistency of the
ansatz for the fermionic sector has been checked for the reduction of the $%
N=1$ ten dimensional supregravity on $S^{3}\times S^{3}$ to four dimensions
\cite{CV}. The reduction considered here is very similar to that case and
the consistency check should follow the same steps.


\vskip 2cm

{\bf Acknowledgments} \medskip W. Sabra would like to thank N. N. Khuri and
his group for hospitality at Rockefeller University where most of this work
was done.

\end{document}